\documentclass[aps,prl,twocolumn,superscriptaddress,showpacs,amsmath,amssymb]{revtex4}
\usepackage[utf8]{inputenc} \usepackage{amsmath,amsthm}
\usepackage{dsfont} \usepackage{graphicx} \usepackage[colorlinks =
true, linkcolor = blue, urlcolor = blue, citecolor = blue, anchorcolor
= blue]{hyperref}
\usepackage{braket} \usepackage{xcolor}
\allowdisplaybreaks

\definecolor{myred}{RGB}{0,0,0}
\newcommand{\rev}[1]{{\color{myred}#1}}

\global\long\def\tr{\mathop{\mathrm{tr}}\limits}
\global\long\def\bra#1{\left\langle#1\right|}
\global\long\def\ket#1{\left|#1\right\rangle}

\global\long\def\al{\alpha} 
\global\long\def\ga{\gamma} 
\global\long\def\De{\Delta} \global\long\def\Ga{\Gamma}

\global\long\def\la{\lambda} 
\global\long\def\si{\sigma}

\global\long\def\transpose#1{#1^\mathrm{T}}
\global\long\def\vec#1{\mathrm{vec}\left(#1\right)}
\def\RE{\mathop{\rm Re}} \def\IM{\mathop{\rm Im}}

\begin{document}

\title{Manifolds of exceptional points and effective Zeno limit of an
  open two-qubit system}

\author{Vladislav Popkov} \affiliation{Faculty of Mathematics and
  Physics, University of Ljubljana, Jadranska 19, SI-1000 Ljubljana,
  Slovenia} \affiliation{Bergisches Universit\"at Wuppertal, Gauss
  Str. 20, D-42097 Wuppertal, Germany} \author{Carlo Presilla}
\affiliation{Dipartimento di Matematica, Sapienza Universit\`a di
  Roma, Piazzale A. Moro 2, 00185 Rome, Italy} \affiliation{Istituto
  Nazionale di Fisica Nucleare, Sezione di Roma 1, 00185 Rome, Italy}
\author{Mario Salerno} \affiliation{Dipartimento di Fisica
  "E.R. Caianiello", Universit\`a di Salerno, Via Giovanni Paolo II,
  84084 Fisciano (SA), Italy}

\pacs{a1,b2} \date{\today}

\begin{abstract}
  \rev{ We analytically investigate the Liouvillian exceptional point
    manifolds (LEPMs) of a two-qubit open system, where one qubit is
    coupled to a dissipative polarization bath. Exploiting a $Z_2$
    symmetry, we block-diagonalize the Liouvillian and show that one
    symmetry block yields two planar LEPMs while the other one
    exhibits a more intricate, multi-sheet topology. The intersection
    curves of these manifolds provide a phase diagram for effective
    Zeno transitions at small dissipation.  These results are
    consistent with a perturbative extrapolation from the strong Zeno
    regime. Interestingly, we find that the fastest relaxation to the
    non-equilibrium steady state occurs on LEPMs associated with the
    \rev{transition to the} effective Zeno regime. }
\end{abstract}
\maketitle

\textit{Introduction.}---%
Strong spectral response to perturbations occurs in open quantum
systems at their branch-point singularities, the so-called exceptional
points (EPs)~\cite{Katino,EP-Heiss,2019ScienceEP}. This phenomenon can
be used for sensing~\cite{2023WiersigReviewSensing,2023EPapplied},
hardware encryption~\cite{2023NatureEncription}, optimizing
performance of quantum thermal machines~\cite{2022NatureZhang},
realizing a multipoint switch between modes in photonic
systems~\cite{2023NatureSwitch}, and other
applications~\cite{2023ReviewEP}.

In classical and semiclassical systems that ignore quantum jumps,
\rev{EPs are typically associated with degeneracies of non-Hermitian
  Hamiltonians, and many theoretical aspects are well understood.}  At
the quantum level, \rev{EPs appear in Markovian open systems, i.e.,
  quantum dynamical semigroups whose time evolution obeys a Lindblad
  master equation, with a time-independent Liouvillian super-operator
  as generator~\cite{Lindblad,GKS,BP}.  In these systems, exceptional
  points of the Liouvillian (LEPs) occur in the parameter space where
  Liouvillian eigenvalues and eigenvectors coalesce.}

In contrast to the EPs of non-Hermitian Hamiltonians, the LEPs include
quantum jumps that reflect the open nature of the systems and allow a
comprehensive understanding of their dynamics in interaction with the
environment~\cite{2019-Minganti}. Moreover, LEPs provide information
that is crucial in the analysis of rapidly decaying states in systems
subject to decoherence~\cite{2022-NewJPhys-Xu}.

The distribution of LEPs driven by the interplay between non-Hermitian
dynamics and dissipation gives rise to Liouvillian exceptional point
manifolds (LEPMs) in the parameter space.  \rev{Understanding these
  manifolds is crucial for controlling the system, such as optimizing
  sensing applications near LEPs by tuning system parameters.
  Conversely, operating in regions where LEPs do not occur is
  essential for applications that require stability, such as quantum
  computing~\cite{2024-Molitor}.  Additionally, the knowledge of LEPMs
  helps to identify parameter regions where the transition to the Zeno
  regime can be achieved with minimal dissipative coupling to the
  environment.}
\begin{figure}[t]
  \centerline{ \includegraphics[width=1\columnwidth,clip]{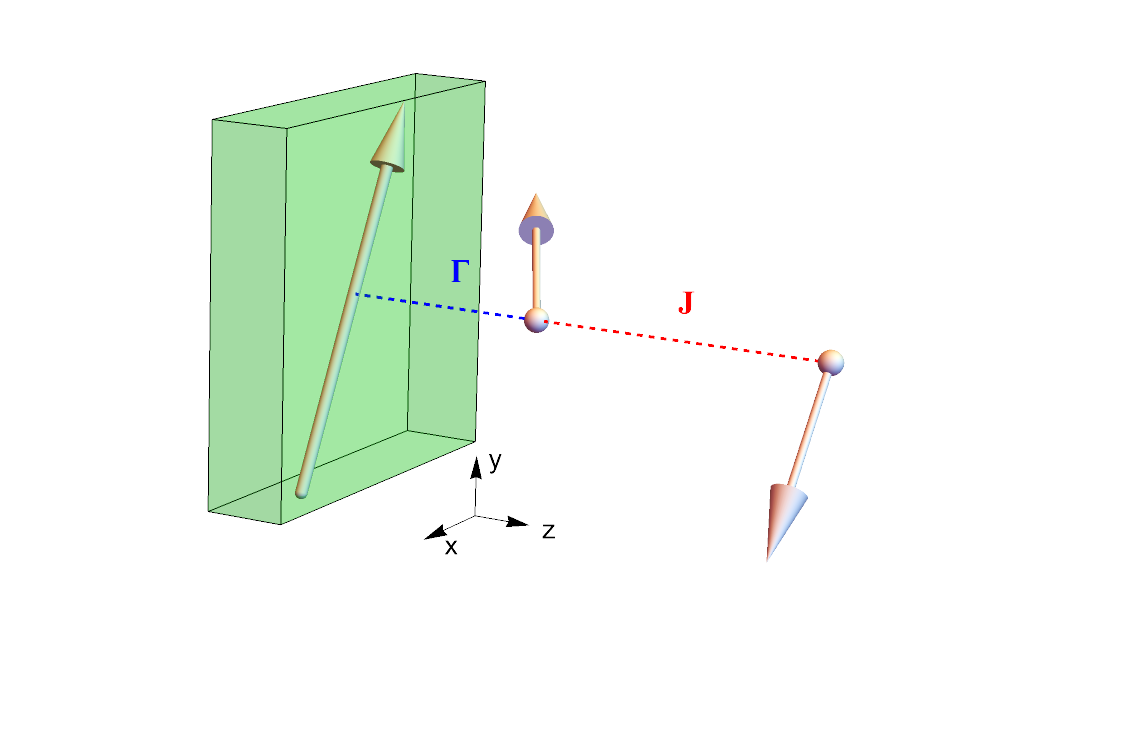}}
  \vskip -1.0cm
  \caption{ An open two-qubit system is schematized as two $XYZ$
    Heisenberg spins $1/2$ interacting via anisotropic exchange
    energies of strength $J\equiv(J_x,J_y,J_z)$. The system is coupled
    to a polarization dissipative bath (green box with a large arrow
    inside) only through one of the two spins. The red and blue dashed
    lines represent the Heisenberg exchange interaction between the
    two spins and the coupling $\Gamma$ between the first spin and the
    bath, respectively. Small spheres indicate the sites on which the
    spins are located. }
  \label{SystemFig}
\end{figure}

Except for a few solvable cases involving two~\cite{N-Seshadri2024}
and three-level systems~\cite{2023-3level}, LEPMs, to our knowledge,
are practically unexplored~\cite{PRL-Ueda-2021} (we discard all cases
for which LEPs reduce to EPs of non Hermitian Hamiltonians as for
example in \cite{PRL-Ueda-2021}).

The aim of the present letter is to provide a full analytical
characterization of all the LEPMs of a two-qubit system and to show
how LEPMs can be used to optimize the phase transitions of the system
to an effective Zeno regime by keeping the dissipative couplings with
the environment as small as possible.

In particular, we consider two $XYZ$ Heisenberg spins $1/2$
interacting with exchange anisotropies $J\equiv(J_x, J_y, J_z)$ and
coupled to a dissipative polarization bath through one of the two
spins only \rev{(see Fig.~\ref{SystemFig} for a scheme and
  \cite{2016-ExperimetalXYZ,2022-ExperimetalXYZ,2021-ExperimentalBath,
    2013-ExperimentalBath} for experimental realizations of a spin
  chain as well as of a polarization bath)}.
The parameter space is three-dimensional and consists, without loss of
generality, of the two parameters, $\gamma=J_y/J_x, \Delta=J_z/J_x$
(\rev{i.e., we work in units of $J_x=1$}), for the Hamiltonian and one
parameter, $\Gamma$, fixing the strength of the dissipative coupling
to the bath.  We take advantage of a $Z_2$ symmetry to
block-diagonalize the Liouvillian super-operator into two blocks, one
of which is independent of $\Delta$.

Using this symmetry, we derive polynomial equations that describe all
the LEPMs. We show that, by restricting only to real values of the
parameters, for the $\Delta$-independent block, LEPMs reduce to two
planes, $\Gamma=8$ and $\Gamma=8 \gamma$, while for the other block,
they exhibit more intricate topologies with a number of sheets
(branches) varying between $1$ and $6$, depending on parameter values.

Quite remarkably, from the intersection curves of some of these
surfaces, it is possible to derive a phase diagram in the
$(\gamma,\Delta)$ plane for effective Zeno transitions occurring at
small dissipation.  \rev{Also notable is the fact that the fastest
  relaxation to the NESS from an initial perturbation occurs precisely
  at the LEPMs that lie at the boundary with the effective Zeno
  regime.} These results are in agreement with a perturbative
calculation that extrapolates the strong Zeno regime to small
dissipations.

\textit{Two qubit model and Liouvillian symmetry.}---%
We consider an open system of two qubits undergoing an anisotropic
exchange interaction of strength $(1, \gamma, \Delta)$ in the
$(x, y, z)$ directions, respectively, and a Markovian dissipation of
strength $\Gamma$ acting only on qubit 1. The reduced density matrix
$\rho$ of the system evolves in time according to the Lindblad master
equation
\begin{align}
  \frac{\partial \rho}{\partial t} = \mathcal{L} \rho
  \,\, \equiv  -i [H,\rho] + \Ga\left(L \rho L^\dag
  - \tfrac12 (L^\dag L \rho + \rho L^\dag L) \right),
  \label{LME}
\end{align}
with jump operator $L=\sigma^+ \otimes I_2$, $I_n$ being the
$n\times n$ identity matrix, and Hamiltonian
\begin{align}
  H = \si^x \otimes \si^x + \ga \, \si^y \otimes \si^y
  + \De \, \si^z \otimes \si^z.
\end{align}
As usual, $\sigma^\pm = (\sigma^x \pm i \sigma^y)/2$, with
$\sigma^x, \sigma^y, \sigma^z$ being the Pauli matrices.  \rev{Besides
  $J_x=1$, we work also in units of $\hbar=1$ so that in
  Eq.~\eqref{LME} we have} $t=t_{ph}J_x/\hbar$ and
$\Gamma=\Gamma_{ph}\hbar/J_x$, where $t_{ph}$ and $\Gamma_{ph}$ are
the physical time and the physical dissipation strength obtained for
the effective values of $\hbar$ and $J_x$. In present units the
parameters $\gamma,\Delta,\Gamma$ as well as the time $t$ and the
operators $H$ and $\mathcal{L}$ are dimensionless. Note also that
$H=H(\gamma ,\Delta)$ and
$\mathcal{L}=\mathcal{L}(\gamma, \Delta, \Gamma)$.  For $H=0$, the
dissipative term proportional to $\Ga$ would result in relaxation of
the first spin into the fully polarized state in the $z$ direction,
namely $\ket{\uparrow}\bra{\uparrow}$, where
$\sigma^z \ket{\uparrow}=\ket{\uparrow}$, with a relaxation time of
order $1/\Ga$. We will find the solution of Eq.~(\ref{LME}) by solving
the associated eigenvalue problem $\mathcal{L} \rho = \lambda \rho$ in
vector form~\cite{note_vec}
$\vec{\mathcal{L} \rho} = \mathcal{L}_\mathrm{vec} \vec{\rho} =
\lambda \,\vec{\rho}$.  The corresponding vectorized Liouvillian is
given by the $16 \times 16$ matrix
\begin{align}
  \label{liouv}
  \mathcal{L}_\mathrm{vec}
  =& -i H\otimes I_4  + i I_4 \otimes \transpose{H}
  \\
   &+ \Gamma  \left(
     L \otimes L^*
     -\tfrac12 (L^\dagger L) \otimes I_4
     -\tfrac12 I_4 \otimes \transpose{(L^\dagger L)}
     \right). \nonumber
\end{align}
Both the Hamiltonian $H$ and the Lindblad operator $L$ commute with
the operator $\Sigma_z = \sigma^z \otimes \sigma^z$, namely,
\begin{align} [\Sigma_z, H] = 0, \qquad [\Sigma_z, L] = 0.
\end{align}
These relations can be used to block-diagonalize the vectorized
Liouvillian $\mathcal{L}_\mathrm{vec}$ as follows.  Introduce the
matrices
$Q_{\pm} = \frac{1}{2} (I_{16} \pm \Sigma_z \otimes \Sigma_z )$ which
satisfy
\begin{align}
  \begin{split}
    &Q_{+} + Q_{-} = I_{16}, \qquad [Q_{\pm}, \mathcal{L}_\mathrm{vec}] = 0,
    \\
    &Q_{\pm} Q_{\mp} = 0, \qquad \qquad
      Q_{\pm} \mathcal{L}_\mathrm{vec} Q_{\mp}=0.
  \end{split}
  \label{ort}
\end{align}
From these relations we have
$\mathcal{L}_\mathrm{vec}=\mathcal{L}_{+} + \mathcal{L}_{-}$, where
$\mathcal{L}_{\pm}=Q_{\pm}\mathcal{L}_\mathrm{vec} Q_{\pm}$ are
matrices of rank 8 satisfying $\mathcal{L}_{\pm}
\mathcal{L}_{\mp}=0$. The block diagonalization of the vectorized
Liouvillian is then achieved as
\begin{align}
  \mathcal{L}_\mathrm{vec}=\Sigma_+ \oplus \Sigma_- = \left(
  \begin{array}{cc}
    \Sigma_+ & 0 \\
    0 & \Sigma_- \\
  \end{array}
  \right),
\end{align}
where $\Sigma_\pm$ are $8 \times 8$ diagonal-blocks, obtained from
$\mathcal{L}_{\mp}=0$ by eliminating the eight null rows and columns
present in these matrices, see \cite{SM} for explicit matrix elements.

\textit{Liouvillian spectrum and LEPMs.}---%
The full Liouvillian spectrum can be obtained by diagonalizing the
blocks $\Sigma_\pm$ separately. Since the parameter $\Delta$ appears
only in the lower diagonal block $\Sigma_-$~\cite{SM}, the block
diagonalization allows to separate the $\Delta$-dependent and
$\Delta$-independent parts of the spectrum into two orthogonal spaces.

The secular equation for the eigenvalues of the block $\Sigma_+$ is
given by
$\lambda (\Gamma +\lambda ) (\Gamma +2 \lambda )^2
\Lambda(\gamma,\Gamma)=0$, where $\Lambda(\gamma,\Gamma)$ is the
quartic polynomial in $\lambda$
\begin{align*}
  &\Lambda(\gamma,\Gamma)
    = \lambda^4 + 2 \Gamma \lambda^3 +
    \left[8(1+\gamma^2)+ \frac{5}{4}\Gamma^2\right] \lambda^2
  \\
  &+\left[8 (1+\gamma^2)\Gamma+\frac {\Gamma^3}{4}\right] \lambda
    +2\left[8(1+\gamma^4)+\Gamma^2+\gamma^2 (\Gamma^2-16)\right].
\end{align*}
Thus, the eight $\Delta$-independent eigenvalues of $\Sigma_+$ are
\begin{align}
  &\lambda(\gamma, \Gamma)
    = \bigg\{
    \, 0, \, -\Gamma, \, -\frac{\Gamma}{2}, \, -\frac{\Gamma}{2}, \,
    -\frac{\Gamma}{2} \pm\frac{\sqrt{2}}{4}
    \nonumber
  \\&~\times
  \sqrt{ \Gamma^2 -32 (1+\gamma^2)
  \pm \sqrt{(\Gamma^2-64) (\Gamma^2-64\gamma^2)} }~ \bigg\}.
  \label{delta-indep-eig}
\end{align}
\begin{figure}
  \centering \includegraphics[width=0.9\columnwidth,clip]{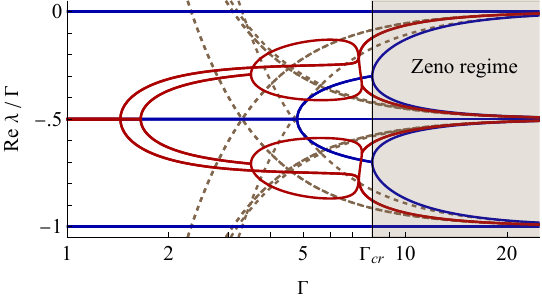}
  \includegraphics[width=0.9\columnwidth,clip]{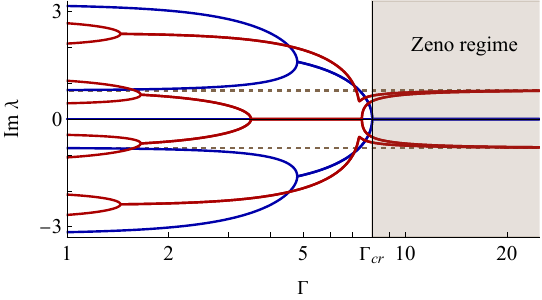}
  \caption{\rev{Real (top panel) and imaginary (bottom panel) parts of
      the eigenvalues $\lambda$ of the Liouvillian operator
      (\ref{liouv}) versus $\Ga$. Red and blue lines refer to
      eigenvalues of the $\Sigma_+$ and $\Sigma_-$ blocks of
      $\mathcal{L}_\mathrm{vec}$, respectively. LEPs coincide with the
      eigenvalue branching points and correspond to broken/resumed
      symmetries (change of eigenvalue degeneracies). Dashed lines
      denote perturbative extrapolations of the eigenvalues from the
      strong, $\Ga \gg \Gamma_{cr}$, Zeno regime, to the low
      dissipation Zeno regime, $\Ga \approx \Gamma_{cr}$, with
      $\Gamma_{cr}$ given in (\ref{GaCrit}). Note the logarithmic
      horizontal scale and the rescaling of $\RE \lambda$ by $\Gamma$
      operated in the top panel.} Parameters are fixed as: $\ga =0.6$,
    $\De=0.4$, corresponding to point b in the bottom panel of
    Fig.~\ref{Fig-PhaseDiagram}.  }
  \label{FigPar1}
\end{figure}
From Eq.~(\ref{delta-indep-eig}) we see that these eigenvalues always
have branching points at two different values of $|\Gamma|$ namely,
$|\Ga| =8$ and $|\Ga|=8 |\ga|$. \rev{Apart from these,
  $\lambda(\gamma,\Gamma)$ has no other singular points.}  The
\rev{remaining} eight $\De$-dependent eigenvalues of the block
$\Sigma_-$ are given by, see \cite{SM} for details,
\begin{align}
  \lambda  =
  \frac{1}{2}\left( -\Gamma \pm \sqrt{\Gamma^2 + \xi_i} \right),
  \qquad i=1,\dots,4,
  \label{delta-dep-eig}
\end{align}
where $\xi_i=\xi_i(\gamma,\Delta,\Gamma)$ are the roots of the quartic
polynomial
\begin{equation}
  c_a \xi^4 +c_b \xi^3 +c_c \xi^2 + c_d \xi+c_e=0,
  \label{polyroots}
\end{equation}
with coefficients $c_a,c_b,c_c,c_d,c_e$, which depend on
$\ga,\De,\Ga$, \rev{given in} \cite{SM}.  \rev{ The branching points
  of the $\Sigma_-$ eigenvalues are obtained by equating to zero the
  discriminant of the polynomial in Eq.~\eqref{polyroots}, this
  leading to the following } eight degree polynomial equation in the
$Z=\Gamma^2$ variable
\begin{equation}
  \sum_{i=0}^8  a_i(X,Y) Z^i = 0,
  \label{polyLEP}
\end{equation}
with coefficients $a_i$ that are polynomials in $Y=\Delta^2$ with
coefficients which are polynomials in $X=\gamma^2$ \cite{SM}.

We stress that the equations determining the branching points of both
the $\Sigma_+$ and $\Sigma_-$ eigenvalues depend on the squares of the
parameters $\gamma,\Delta,\Gamma$. It follows that our results are
invariant with respect to a change of sign of each one of these
parameters. Hereafter, for simplicity, we will assume that
$\gamma,\Delta,\Gamma$ are all positive~\cite{nota_reali}.  Depending
on the values of the parameters $\ga$ and $\De$, we have LEPs,
corresponding to branching points in $\RE \la(\Ga)$ and $\IM \la(\Ga)$
at up to $2+6=8$ different values of $\Gamma$.  An example with LEPs
at 2+5 different values of $\Gamma$ is shown in Fig.~\ref{FigPar1},
other examples are given in \cite{SM}.
\begin{figure}
  \centering \includegraphics[width=0.8\columnwidth,clip]{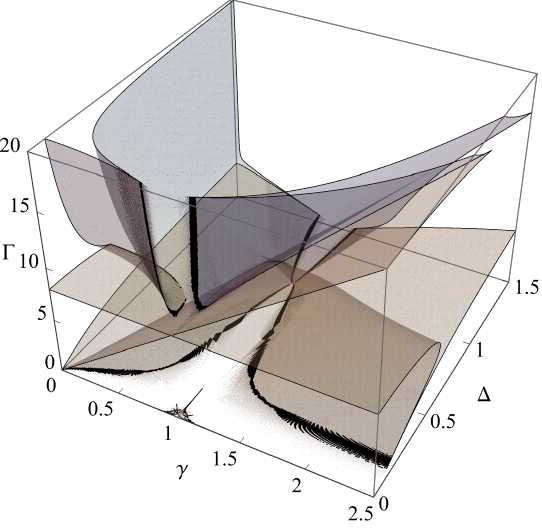}
  \caption{ Two-dimensional LEP manifolds of the two-qubit open system
    (\ref{LME}) in the three-dimensional parameter space
    $\{\gamma, \Delta, \Gamma\}$. Only the LEP manifolds originating
    from the $\Delta$-dependent block $\Sigma_-$ of the Liouvillian
    are shown here. }
  \label{Fig-3D}
\end{figure}
In general, at an EP the eigenvalue coalescence goes along with an
eigenvector coalescence~\cite{Kanki2017}, this means non
diagonalizability of the Liouvillian at any LEP. In \cite{SM} we
provide analytic expressions of the Liouvillian eigenvectors on
various LEPMs explicitly showing the characteristic Jordan block
decomposition of $\Sigma_\pm$.

In the three-dimensional space of parameters $\ga,\De,\Ga$ the values
of $\Gamma$ for which LEPs are found form two-dimensional manifolds
$\Gamma(\gamma,\Delta)$. The LEP manifolds originating from the
$\Delta$-dependent block $\Sigma_-$ of the Liouvillian are shown in
Fig.~\ref{Fig-3D}. Due to the divergence of some manifolds in the
limit $\Delta\to 0$, see later, the plot is limited to
$\Delta\geq 0.05$.  The two manifolds originating from the
$\Delta$-independent block $\Sigma_+$ of the Liouvillian, not shown in
Fig.~\ref{Fig-3D}, correspond to the straight planes $\Ga=8$ and
$\Ga=8|\ga|$.
\begin{figure}
  \centering \includegraphics[width=0.75\columnwidth,clip]{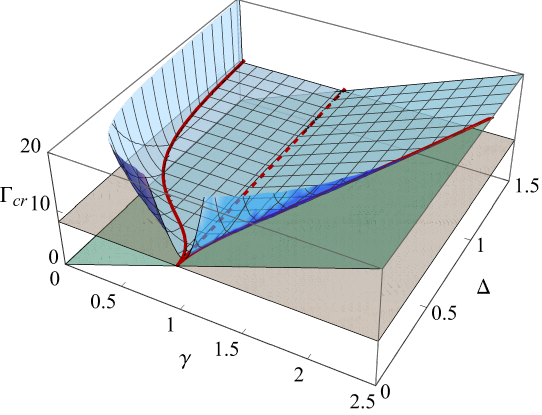}
  \hspace{0.1\columnwidth}
  \includegraphics[width=0.75\columnwidth,clip]{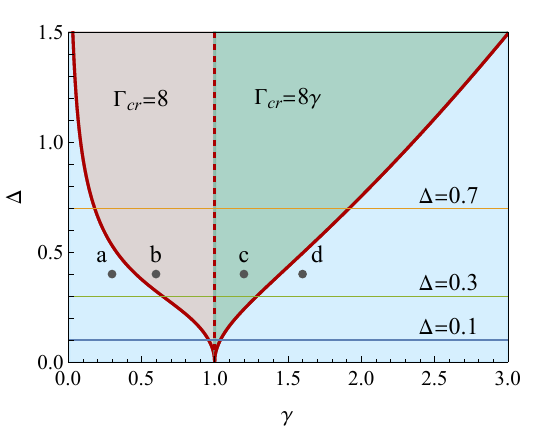}
  \caption{{\bf Top panel.} Behavior of $\Ga_{cr}(\gamma,\Delta)$,
    light-blue surface with a mesh, for $\Delta \ge 0.05$.
    \rev{Within} the region delimited by the red continuous line this
    surface coincides with the straight planes $\Gamma_{cr}=8$
    (displayed in gray) and $\Gamma_{cr}=8\gamma$ (displayed in
    green).  {\bf Bottom panel.}  Two-dimensional diagram of
    $\Ga_{cr}(\gamma,\Delta)$.  The detailed branching point structure
    of the Liouvillian eigenvalues at the point labelled b is shown in
    Fig.~\ref{FigPar1}. The same info at points a, c, and d is
    provided in \cite{SM}. In \cite{SM} we also show the sections of
    $\Ga_{cr}(\gamma,\Delta)$ at the planes $\Delta=0.1,0.3,$ and
    $0.7$ whose projections are indicated here by tiny solid lines.  }
  \label{Fig-PhaseDiagram}
\end{figure}

Aiming at Zeno limit applications, we examine the location of the LEPs
with the largest $\Ga$ value. Denoting by $\Ga_{\mathrm{LEP},j}$ the
$\Ga$-coordinates of the branching points in the $(\Ga, \la_j)$ plane
(see Fig.~\ref{FigPar1}), we define $\Ga_{cr}$ as the $\Gamma-$value
beyond which all eigenvalues are analytical:
\begin{align}
  \Ga_{cr}(\ga,\De)  =  \sup_j \Ga_{\mathrm{LEP},j},
  \label{GaCrit}
\end{align}
\rev{ The behavior of $\Ga_{cr}(\ga,\De)$ is shown in
  Fig.~\ref{Fig-PhaseDiagram} by a bare three-dimensional plot in the
  top panel and a $(\gamma,\Delta)$ diagram in the bottom
  panel. Remarkably, the $(\gamma,\Delta)$ plane is divided into two
  regions by a solid red boundary line (given parametrically in
  \cite{SM}). The region above this line is further split by the
  dashed line $\gamma=1$ into two parts: the gray region $\ga<1$ where
  $\Gamma_{cr}=8$, and the green region $\ga>1$ where
  $\Gamma_{cr}=8\gamma$. In the blue region below the solid red line,
  $\Gamma_{cr}$ corresponds to LEPs arising from eigenvalues of
  $\Sigma_-$ and depends on both $\gamma$ and $\Delta$. In this region
  $\Gamma_{cr}$ increases as $\Delta$ is decreased, and diverges (see
  Eq.~(\ref{eq:GaDivergencies})) as $\Delta\to 0$, except at
  $\gamma=1$; the same holds for $\gamma \rightarrow 0$. In contrast,
  in the gray and green regions of the phase diagram, no singularities
  occur and the dynamics becomes fully analytic. This analyticity
  allows for an expansion in $1/\Gamma$, leading to an effective
  near-Zeno dynamics, as explained below.}
\begin{figure}
  \centerline{\includegraphics[width=0.48\columnwidth,clip]{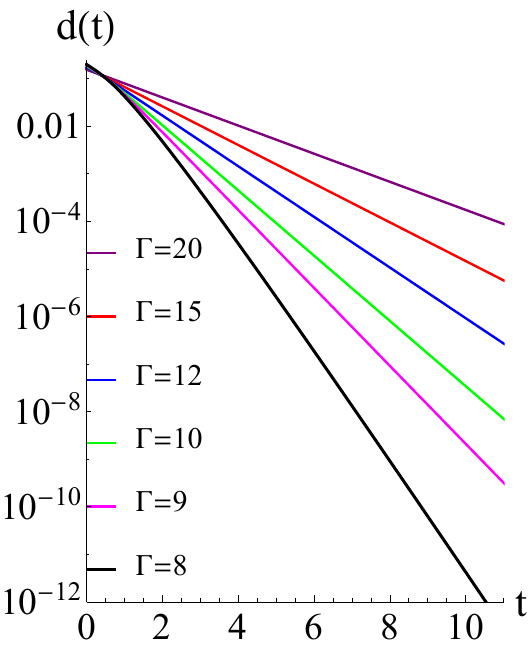}
    \includegraphics[width=0.48\columnwidth,clip]{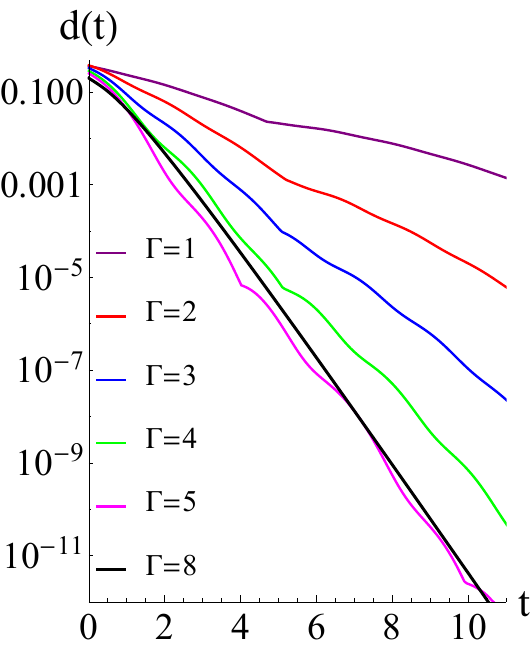} }
  \caption{\rev{Distance
      $d(t)\equiv {\|\rho(t)-\rho_{\text{NESS}}(\gamma_f)\|}_2$ versus
      time $t$ for rapid quenches in the parameter $\gamma$, from
      $\gamma_{i}=0.4$ to $\gamma_{f}=0.8$, on different
      constant-$\Gamma$ planes as indicated in the legend, for
      arbitrary $\Delta$. The left panel shows data for planes above
      the $\Gamma=8$ LEPM, while the right panel data for planes below
      it.  }}
  \label{Gamma8-Relax}
\end{figure}

\rev{ \textit{Relaxation times near and on LEPMs.}---%
  Interestingly, we find that in the region of parameter space
  characterized by $\Gamma_{cr}=8$ (see Fig.~\ref{Fig-PhaseDiagram})
  the fastest relaxation from an initial perturbed state to the exact
  NESS given in \cite{SM} is achieved on the $\Gamma=8$ LEPM. To show
  this we perform instantaneous quenches of the parameter \(\gamma\)
  from \(\gamma_i\) to \(\gamma_f\) on different constant-\(\Gamma\)
  planes, keeping \(\Delta\) fixed.  The quench is implemented by
  setting the initial condition as
  \(\rho_i = \rho_{\text{NESS}}(\gamma_i, \Gamma)\), the exact NESS
  (see Eq.~(\ref{rhoness}) in~\cite{SM}) corresponding to
  \(\gamma_i\), for the time evolution governed by Eq.~(\ref{LME})
  with \(\mathcal{L}\) corresponding to \(\gamma_f\), i.e.,
  \(\mathcal{L} \equiv \mathcal{L}(\gamma_f, \Delta, \Gamma)\). To
  characterize the relaxation dynamics, we compute the distance
  between \(\rho(t)\) and $\rho_{\text{NESS}}(\gamma_f)$, and plot it
  as a function of time. The results are shown in
  Fig.~\ref{Gamma8-Relax}.}

\rev{From the left panel, we observe that the relaxation time
  increases above $\Gamma = 8$. This slower relaxation is due to the
  onset of the quantum Zeno regime where the effective dissipation
  strength starts to decrease with $\Gamma$ (see Eq. (\ref{LMEeff})).
  The relaxation curves in this regime are smooth and uniform, as
  expected due to the absence of LEPMs above $\Gamma=8$.  In contrast,
  below the LEPM plane, right panel, the relaxation curves develop
  undulations or cusps, which we attribute to interference with other
  LEPM branches present below the $\Gamma = 8$ plane.  Just below the
  LEPM plane, the relaxation time decreases slightly before rapidly
  increasing as $\Gamma$ is reduced further.  A similar behavior is
  observed when $\Gamma_{cr}=8\gamma$ (see~Fig.~\ref{8gamma-relax}
  in~\cite{SM}).  }

\textit{Effective near Zeno dynamics.}---%
If the dissipation is strong, and in the absence of branching points,
i.e., for $\Ga \gg \Ga_{cr}$, all Liouvillian eigenvalues can be
calculated explicitly using a perturbative Dyson
series~\cite{2021ZenoSpectrumPRL}.  The complete set of $16$
Liouvillian eigenvalues up to order $1/\Ga$ included is given by
\begin{align*}
  \begin{split}
    \la_{0,\al}
    &= \left\{ 0,\  - 2\frac { \ga_{+}} {\Ga},\
      - \frac { \ga_{+}} {\Ga}  \pm 2 \De i  \right\},
    \\
    \la_{1,\al}
    &= \left\{ - \frac{\Ga}{2},\
      - \frac{\Ga}{2},\ - \frac{\Ga}{2}  \pm  \frac{2 \ga_{-}}{\Ga},\
      - \frac{\Ga}{2} \pm \frac{8 \ga}{\Ga}   \pm 2  \De  i \right\},
    \\
    \la_{2,\al}
    &= \left\{ -\Ga,\  -\Ga  + 2\frac { \ga_{+}} {\Ga},\
      -\Ga  + \frac { \ga_{+}} {\Ga} \pm 2 \De i \right\},
  \end{split}
\end{align*}
where $\ga_{\pm}= 4(1\pm \ga^2)$.  The eigenvalues are labeled by a
stripe index, numbers 0,1,2, and a mode index, Greek letters $\al$
ranging from 1 to 4 for stripes 0 and 2 and from 1 to 8 for stripe 1,
see \cite{2021ZenoSpectrumPRL}.  The modes $\la_{0,\al}$ contain the
nonequilibrium stationary state (NESS), corresponding to the null
eigenvalue, and the slowest relaxation modes, which determine the late
time evolution.  On the other hand, the modes
$\la_{1, \al},\la_{2, \al}$ have large negative real parts and thus
correspond to rapid relaxation of the first spin towards the target
state, namely, the qubit fully polarized along the $z$-axis.  The
\rev{above asymptotic expressions for Liouvillian eigenvalues are
  shown in} Fig.~\ref{FigPar1} by dashed lines.

The theory also predicts~\cite{2021ZenoSpectrumPRL} explicit analytic
expressions for the Liouvillian eigenvectors as well as an effective
near Zeno dynamics.  Namely, while the first spin relaxes towards the
target state $\ket{\uparrow}\bra{\uparrow}$ in a time $t=O(1/\Ga)$,
the second spin has a slower dynamics described by a reduced effective
Lindblad equation \cite{2018ZenoDynamics}.  In fact, for
$t\gg 1/\Gamma$ up to an error $O(1/\Gamma^2)$ we have
$\rho(t)=\ket{\uparrow}\bra{\uparrow} \otimes R(t)$ with $R(t)$
satisfying
\begin{align}
  \frac{\partial R}{\partial t}
  &=-i [h_D,R] +\frac{1}{\Ga}
    (
    \tilde{L} R \tilde{L}^\dagger -
    \tfrac12 ( \tilde{L}^\dagger \tilde{L} R
    + R \tilde{L}^\dagger \tilde{L}) ),
    \label{LMEeff}
\end{align}
where $h_D=\De \si^z$ and $\tilde{L}=4(\si^x + i \ga \si^y)$.  Note
that for $R(t)$, \rev{i.e., for the density matrix of the spin not
  directly affected by the dissipation, the effective dissipation
  strength is} $1/\Gamma$ and not $\Gamma$ as in Eq.~\eqref{LME} for
$\rho(t)$.  The near Zeno limit NESS is found straightforwardly as the
time independent solution of Eq.~(\ref{LMEeff}). This yields
\begin{align}
  \rho_\mathrm{Zeno}= \ket{\uparrow} \bra{\uparrow}\otimes
  \left(
  \begin{array}{cc}
    \frac{(\gamma +1)^2}{2 \left(\gamma^2+1\right)} & 0 \\
    0 & \frac{(\gamma -1)^2}{2 \left(\gamma^2+1\right)} \\
  \end{array}
  \right) + O\left( \frac{1}{\Ga} \right). \label{eq:ZenoNESS}
\end{align}

\rev{To apply the near Zeno limit predictions, we must ensure that the
  $1/\Gamma$ Dyson perturbative expansion is convergent. Convergent
  series, which yield a unique sum, fail across branching
  points. Therefore, the $1/\Ga$ Taylor series for each Liouvillian
  eigenvalue $\la_j$ is expected to have a convergence radius of
  $1/\Ga_{\mathrm{LEP}}(\la_j)$, where $\Ga_{\mathrm{LEP}}(\la_j)$ is
  the largest value of $\Gamma$ at which a branching point occurs (see
  Fig.~\ref{FigPar1}). Beyond the critical point $\Ga_{cr}$ defined
  above, \textit{all} Liouvillian eigenvalues become analytic. On the
  $(\ga,\De)$ plane, $\Ga_{cr}$ is finite except near $\De=0$ and
  $\ga=0$, where singularities appear.  } We find that $\Ga_{cr}(\De)$
for $\De \rightarrow 0^+$ behaves as
\begin{align}
  \Ga_{cr}(\De\rightarrow 0^+) = \max \left(
  \left| \frac{2(\gamma^2-1)}{\Delta} \right|,
  \left| \frac{ \gamma^3- 1/\gamma }{\Delta}    \right| \right).
  \label{eq:GaDivergencies}
\end{align}
This singularity originates from the fact that at $\De=0$ the spectrum
of the dissipation projected Hamiltonian $h_D=\De \si^z$ becomes
degenerate.

\textit{Conclusions.}- We analytically investigated the Liouvillian
spectrum of two qubit systems and identified regions of the LEPM where
it is possible to achieve an effective Zeno regime at a minimal,
$\Gamma_{cr}$, dissipation.  We provided an analytic description of
the temporal dynamics \rev{which applies in this regime and showed
  that the fastest relaxation time to the NESS occurs precisely on the
  LEPMs characterizing the effective Zeno regime.  The rich LEPM
  structure \rev{uncovered} here could be checked experimentally by
  quantum process tomography, as done in \cite{experimental_LEP}.}

\begin{acknowledgments}
  V.~P.~acknowledges support by ERC Advanced grant No.~101096208 --
  QUEST, and Research Programme P1-0402 of Slovenian Research and
  Innovation Agency (ARIS), and from Deutsche Forschungsgemeinschaft
  through DFG project KL645/20-2.
\end{acknowledgments}

\clearpage
\onecolumngrid
\appendix

\setcounter{page}{1} \setcounter{equation}{0} \setcounter{figure}{0}

\renewcommand{\theequation}{\textsc{S}-\arabic{equation}}
\renewcommand{\thefigure}{\textsc{S}-\arabic{figure}}

\begin{center} {\bfseries \large Supplemental Material:
    \\
    Manifolds of exceptional points and effective Zeno limit of an
    open two-qubit system}
\end{center}

\begin{center}
  Vladislav Popkov,$^{1,2}$ Carlo Presilla,$^{3,4}$ and Mario
  Salerno$^5$
  \\
  $^1$\textit{Faculty of Mathematics and Physics, University of
    Ljubljana, Jadranska 19, SI-1000 Ljubljana, Slovenia}
  \\
  $^2$\textit{{Bergisches Universit\"at Wuppertal, Gauss Str. 20,
      D-42097 Wuppertal, Germany}}
  \\
  $^3$\textit{Dipartimento di Matematica, Sapienza Universit\`a di
    Roma, Piazzale A. Moro 2, 00185 Rome, Italy}
  \\
  $^4$\textit{Istituto Nazionale di Fisica Nucleare, Sezione di Roma
    1, 00185 Rome, Italy}
  \\
  $^5$\textit{Dipartimento di Fisica, Universit\`a di Salerno, Via
    Giovanni Paolo II, 84084 Fisciano (SA), Italy}
\end{center}

This Supplemental Material contains \rev{eight} sections organized as
follows.  In section A we describe the block structure
$\Sigma_+,\Sigma_-$ of the Liouvillian. The polynomial equation whose
roots provide the LEPs arising from $\Sigma_-$ is detailed in section
B.  In C we show some two-dimensional sections of the
three-dimensional LEPMs given in Fig.~\ref{Fig-3D} as well as of
$\Gamma_{cr}(\gamma,\Delta)$ given in Fig.~\ref{Fig-PhaseDiagram}.
The bifurcation diagrams at the four points indicated in
Fig.~\ref{Fig-PhaseDiagram} of the main text is shown in section D.
\rev{In section E we explain how the boundary red line shown in
  Fig.~\ref{Fig-PhaseDiagram} is obtained while in} section F we make
explicit the non-diagonalizability of the Liouvillian on the
LEPMs. \rev{In section G we compare relaxation times to the NESS
  achieved on and out the $\Gamma=8\gamma$ LEPM. Finally, in section G
  we discuss some subtle details regarding quantum Zeno regime which
  did not appear in the main text.}

\section{A. Block diagonalization of the Liouvillian}

\begin{figure*}[htb]
  \centering \includegraphics[width=0.4\columnwidth,clip]{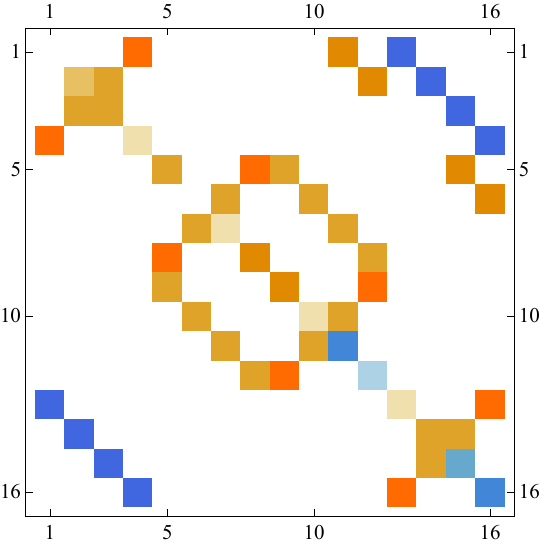}
  \hspace{0.1\columnwidth}
  \includegraphics[width=0.4\columnwidth,clip]{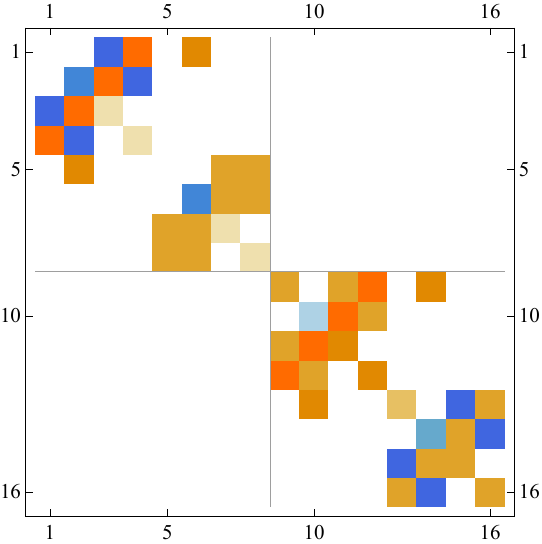}
  \caption{Liouvillian of the dissipative two qubit system as obtained
    from Eq.~\ref{liouv} (left panel) and its block diagonal form
    (right panel) acquired after operating the $\mathcal{U}_\pm$
    transformation. Colors are associated to different matrix
    elements, white regions corresponding to zeros (for other color
    correspondences see Eqs. \ref{Ablock}, \ref{Bblock}). }
  \label{figA1}
\end{figure*}

The $\mathcal{Z}_2$ symmetry discussed in the main text allows to put
the Liouvillian in the form
\begin{equation}
  \mathcal{L}=\left(
    \begin{array}{cc}
      \Sigma_+ & 0 \\
      0 & \Sigma_- \\
    \end{array}
  \right),
\end{equation}
with the two $8 \times 8$ diagonal-blocks, $\Sigma_\pm$, related to
the $\mathcal{Z}_2$ symmetry, achieved by eliminating the eight null
rows and columns from the corresponding $16 \times 16$ matrices
$\Sigma_\pm = \mathcal{U}_\pm \cdot \mathcal{L} \cdot
\mathcal{U}_\pm$. In Fig.~\ref{figA1} we show the block structure of
the Liouvillian $\mathcal{L}$ obtained directly from Eq.~(\ref{liouv})
(left panel) and the one obtained after the $\mathcal{Z}_2$ block
diagonalization discussed in the main text (right panel).

The blocks $\Sigma_{\pm}$ can be further rearranged in the form

\begin{equation}
  \Sigma_+=\left(
    \begin{array}{cc}
      A_1 & C \\
      C & A_2 \\
    \end{array}
  \right),
  \qquad
  \Sigma_-=\left(
    \begin{array}{cc}
      B_1 & C \\
      C & B_2 \\
    \end{array}
  \right)
  \label{Block}
\end{equation}
with $C$ a $4\times4$ matrix whose single nonzero element is
$C_{1,2}=\Gamma$, and $A_1,A_2$ and $B_1,B_2$ are given by:
\begin{align}
  \begin{split}
    A_1 &=
          \begin{pmatrix}
            0 & 0 & -i (1-\gamma ) & i (1-\gamma ) \\
            0 & -\Gamma  & i (1-\gamma ) & -i (1-\gamma ) \\
            -i (1-\gamma ) & i (1-\gamma ) & -\frac{\Gamma }{2} & 0 \\
            i (1-\gamma ) & -i (1-\gamma ) & 0 & -\frac{\Gamma }{2} \\
          \end{pmatrix},
    \\
    \\
    A_2 &=
          \begin{pmatrix}
            0 & 0 & i (\gamma +1) & -i (\gamma +1) \\
            0 & -\Gamma  & -i (\gamma +1) & i (\gamma +1) \\
            i (\gamma +1) & -i (\gamma +1) & -\frac{\Gamma }{2} & 0 \\
            -i (\gamma +1) & i (\gamma +1) & 0 & -\frac{\Gamma }{2} \\
          \end{pmatrix},
  \end{split}
  \label{Ablock}
\end{align}

\begin{align}
  \begin{split}
    B_1
    &=
      \begin{pmatrix}
        2 i \Delta  & 0 & -i (\gamma +1) & i (1-\gamma ) \\
        0 & -\Gamma +2 i \Delta  & i (1-\gamma ) & -i (\gamma +1) \\
        -i (\gamma +1) & i (1-\gamma ) & -\frac{\Gamma }{2}+2 i \Delta  & 0 \\
        i (1-\gamma ) & -i (\gamma +1) & 0 & -\frac{\Gamma }{2}+2 i \Delta  \\
      \end{pmatrix},
    \\
    \\
    B_2
    &=
      \begin{pmatrix}
        -2 i \Delta  & 0 & -i (1-\gamma ) & i (\gamma +1) \\
        0 & -\Gamma -2 i \Delta  & i (\gamma +1) & -i (1-\gamma ) \\
        -i (1-\gamma ) & i (\gamma +1) & -\frac{\Gamma }{2}-2 i \Delta  & 0 \\
        i (\gamma +1) & -i (1-\gamma ) & 0 & -\frac{\Gamma }{2}-2 i \Delta  \\
      \end{pmatrix}.
  \end{split}
  \label{Bblock}
\end{align}
Note that the parameter $\Delta$ appears only in the block $\Sigma_-$.

From the above expressions it is easy to find that the secular
equation for the eigenvalues of the block $\Sigma_+$ provides
\begin{align}
  \lambda (\Gamma +\lambda ) (\Gamma +2 \lambda )^2
  \Lambda(\gamma,\Gamma)=0,
\end{align}
where $\Lambda(\gamma,\Gamma)$ is the quartic polynomial in $\lambda$
\begin{align}
  \Lambda(\gamma,\Gamma)
  = \lambda^4 + 2 \Gamma \lambda^3 +
  \left[8(1+\gamma^2)+ \frac{5}{4}\Gamma^2\right] \lambda^2
  +\left[8 (1+\gamma^2)\Gamma+\frac {\Gamma^3}{4}\right] \lambda
  +2\left[8(1+\gamma^4)+\Gamma^2+\gamma^2 (\Gamma^2-16)\right].
\end{align}
We conclude that the eight $\Delta$-independent eigenvalues of
$\Sigma_+$ are
\begin{align}
  \lambda(\gamma, \Gamma) =
  \left\{
  \, 0, \, -\Gamma, \, -\frac{\Gamma}{2}, \, -\frac{\Gamma}{2}, \,
  -\frac{\Gamma}{2} \pm \frac{\sqrt{2}}{4}
  \sqrt{ \Gamma^2 - 32(1+\gamma^2)
  \pm \sqrt{(\Gamma^2-64) (\Gamma^2-64\gamma^2)} } \right\}.
  \label{SM_delta-indep-eig}
\end{align}

On the other hand, the eight $\De$-dependent eigenvalues of the block
$\Sigma_-$ are given by
\begin{align}
  \lambda =
  \frac{1}{2}\left( -\Gamma \pm \sqrt{\Gamma^2 + \xi_i} \right),
  \qquad i=1,\dots,4,
  \label{SM_delta-dep-eig}
\end{align}
where $\xi_i=\xi_i(\gamma,\Delta,\Gamma)$ are the roots of the quartic
polynomial $c_a \xi^4 +c_b \xi^3 +c_c \xi^2 + c_d \xi+c_e=0 $ with
coefficients
\begin{align}
  \begin{split}
    c_a &= 1,
    \\
    c_b &= 2 \Gamma^2 + 2^5 (1+\gamma^2) +2^6 \Delta^2,
    \\
    c_c &= 2^5 \Gamma^2 \left( 2(1+\gamma^2)+5\Delta^2 \right)
          + \Gamma^4
          + 2^8 \left( 2\gamma^4 + (1+\gamma^2)^2
          + 2(1+\gamma^2)\Delta^2 + 6\Delta^4 \right),
    \\
    c_d &= 2^5 \left( 8\Gamma^2 \left( (1+\gamma^2)^2 + 6\gamma^2
          + 14\Delta^4 \right)
          + \Gamma^4 \left( (1+\gamma^2) + 5\Delta^2 \right) \right.
    \\
        &\quad \left. + 2^8 \left( \gamma^2(1+\gamma^2)
          + \left( (1+\gamma^2)^2 - 6\gamma^2 \right) \Delta^2
          - (1+\gamma^2)\Delta^4
          + 2 \Delta^6 \right) \right),
    \\
    c_e &= 2^6 \Gamma^6 \Delta^2 + 2^{16}
          \left( \gamma^2 - (1+\gamma^2)\Delta^2 + \Delta^4 \right)^2
          + 2^8 \Gamma^4 \left( 4\gamma^2 - 2(1+\gamma^2) \Delta^2
          + 9\Delta^4 \right)
    \\
        &\quad + 2^{12} \Gamma^2 \left( 2 \gamma^2(1+\gamma^2)
          + \left( (1+\gamma^2) -6\gamma^2 \right)\Delta^2
          - 4(1+\gamma^2)\Delta^4 + 6\Delta^6 \right).
  \end{split}
\end{align}
By using Mathematica, we find
\begin{align}
  \begin{split}
    \xi_1 &= -c_b/(4 c_a) - p_4/2 - \sqrt{p_5 - p_6}/2,
    \\
    \xi_2 &= -c_b/(4 c_a) - p_4/2 + \sqrt{p_5 - p_6}/2,
    \\
    \xi_3 &= -c_b/(4 c_a) + p_4/2 - \sqrt{p_5 + p_6}/2,
    \\
    \xi_4 &= -c_b/(4 c_a) - p_4/2 + \sqrt{p_5 + p_6}/2,
  \end{split}
\end{align}
where
\begin{align}
  \begin{split}
    p_1 &= 2 c_c^3 - 9 c_b c_c c_d + 27 c_a c_d^2 + 27 c_b^2 c_e
          - 72 c_a c_c c_e,
    \\
    p_2 &= p_1 + \sqrt{p_1^2 - 4 (c_c^2 - 3 c_b c_d + 12 c_a c_e)^3},
    \\
    p_3 &= (c_c^2 - 3 c_b c_d + 12 c_a c_e)/( 3 c_a \sqrt[3]{p_2/2}) +
          \sqrt[3]{p_2/2}/(3 c_a),
    \\
    p_4 &= \sqrt{c_b^2/(4 c_a^2) - 2 c_c/(3 c_a) + p_3},
    \\
    p_5 &= c_b^2/(2 c_a^2) - 4 c_c/(3 c_a) - p_3,
    \\
    p_6 &= (-c_b^3/c_a^3 + 4 c_b c_c/c_a^2 - 8 c_d/c_a)/(4 p_4).
  \end{split}
\end{align}
Note that with $\sqrt[3]{p_2/2}$ we mean the real-valued cube root of
$p_2/2$.

\section{B. Polynomial equation for the LEP manifolds arising from
  \boldmath{$\Sigma_-$}}
\rev{All Liouvillian exceptional points (LEPs) in the
  ($\gamma,\Delta, \Gamma$) parameter space of the $\Sigma_-$ block
  are determined by requiring that the discriminant of the polynomial
  in Eq. (\ref{polyroots}) vanishes. This condition yields the
  following degree-eight polynomial equation in the
  $Z \equiv \Gamma^2$ variable}
\begin{equation}
  \sum_{i=0}^8  a_i(X,Y) Z^i = 0,
  \label{polyLEP}
\end{equation}
with coefficients $a_i$ given by
\begin{align*}
  a_0 &= 2^{32} ( X - 1)^4 X Y^2 (X^2 + (1 - 4 Y)^2 - 2 X (1 + 4 Y))^2,
  \\
  a_1 &= -2^{27} (X-1)^8 X Y + 2^{31} (X-1)^6 X (X+1) Y^2
        + 2^31 X (X-1)^4(1+30X+X^2)Y^3
  \\
      &\quad- 2^{35}(X-1)^2 X (X+1) (3+2X+3X^2)Y^4 + 2^{36} (X-1)^2 X
        (5 + 6 X + 5 X^2) Y^5 - 2^{38} X (X - 1)^2 ( X + 1 ) Y^6,
  \\
  a_2  &= 2^{20} (-1 + X)^8 X + 2^{20} (-1 + X)^6 (1 + X) (1 - 34 X + X^2) Y
         - 2^{23} (-1 + X)^4 (1 + 24 X + 238 X^2 + 24 X^3 + X^4) Y^2
  \\
      &\quad+ 2^{24} (-1 + X)^2 (1 + X) (1 + 44 X - 602 X^2 + 44 X^3 + X^4) Y^3
        + 2^{29} X (27 + 36 X + 2 X^2 + 36 X^3 + 27 X^4) Y^4
  \\
      &\quad- 2^{33} X (1 + X) (5 - 2 X + 5 X^2) Y^5
        + 2^{33} X (3 + 2 X + 3 X^2) Y^6,
  \\
  a_3 &= -2^{14} (-1 + X)^6 (1 + X) (1 - 18 X + X^2)
        + 2^{18} (-1 + X)^4 (1 + 4 X + 54 X^2 + 4 X^3 + X^4) Y
  \\
      &\quad+ 2^{21} (1 + X) (-1 + X)^2 (1+10 X + 42 X^2 + 10 X^3 + X^4) Y^2
        - 2^{20} (3 + 22 X - 883 X^2 - 332 X^3
  \\
      &\quad- 883 X^4 + 22 X^5 + 3 X^6) Y^3 - 2^{26} X (1 + X) (21 + 22 X
        + 21 X^2) Y^4 + 2^{29} X (5 + 6 X + 5 X^2) Y^5 - 2^{30} X (1 + X) Y^6,
  \\
  a_4 &= -256 (-1 + X)^4 (15 - 60 X - 166 X^2 - 60 X^3 + 15 X^4)
        - 2^{15} (-1 + X)^2 (1 + X) (1 + 4 X - 42 X^2 + 4 X^3 + X^4) Y
  \\
      &\quad- 2^{16} (3 + 42 X - 3 X^2 - 340 X^3 - 3 X^4 + 42 X^5 + 3 X^6) Y^2
        + 2^{16} (1 + X) (3 - 156 X - 974 X^2 - 156 X^3 + 3 X^4) Y^3
  \\
      &\quad+ 2^{21} X (39 + 74 X + 39 X^2) Y^4 - 2^{24}*5 X (1 + X) Y^5
        + 2^24 X Y^6,
  \\
  a_5 &= -64 (3 + X - 21 X^2 + 17 X^3 + 17 X^4 - 21 X^5 + X^6 + 3 X^7)
        - 256 (-1 + X)^2 (1 + 68 X + 246 X^2 + 68 X^3 + X^4) Y
  \\
      &\quad+ 2^{13} (1 + X) (1 - 6 X + X^2) (1 + 14 X + X^2) Y^2 -2^{12}
        (1 - 236 X - 682 X^2 - 236 X^3 + X^4) Y^3 - 2^{18}*9 X (1 + X) Y^4
  \\
      &\quad+ 2^{20} X Y^5,
  \\
  a_6 &= 4 (1 - 2 X - X^2 + 4 X^3 - X^4 - 2 X^5 + X^6)
        + 16 (-1 + X)^2 (1 + X) (5 + 38 X + 5 X^2) Y
  \\
      &\quad- 2^7 (1 - 28 X - 138 X^2 - 28 X^3 + X^4) Y^2
        - 7*2^{12} X (1 + X) Y^3 + 3*2^{13} X Y^4,
  \\
  a_7 &= -(1 + 4 X - 10 X^2 + 4 X^3 + X^4 ) Y
        - 2^7 X ( 1 + X) Y^2 + 2^8 X Y^3,
  \\
  a_8 &=  X Y^2.
\end{align*}
\rev{It is worth noting the multi-polynomial structure of
  Eq.~(\ref{polyLEP}), where the coefficients $a_i$ are polynomials in
  $Y=\Delta^2$, whose coefficients are, in turn, polynomials in
  $X=\gamma^2$. This nested polynomial structure reveals the
  complexity of the LEPMs in the parameter space.  }

\section{C. Two-dimensional cuts of the LEP manifolds}
\begin{figure}[h]
  \centering
  \includegraphics[width=0.42\columnwidth,clip]{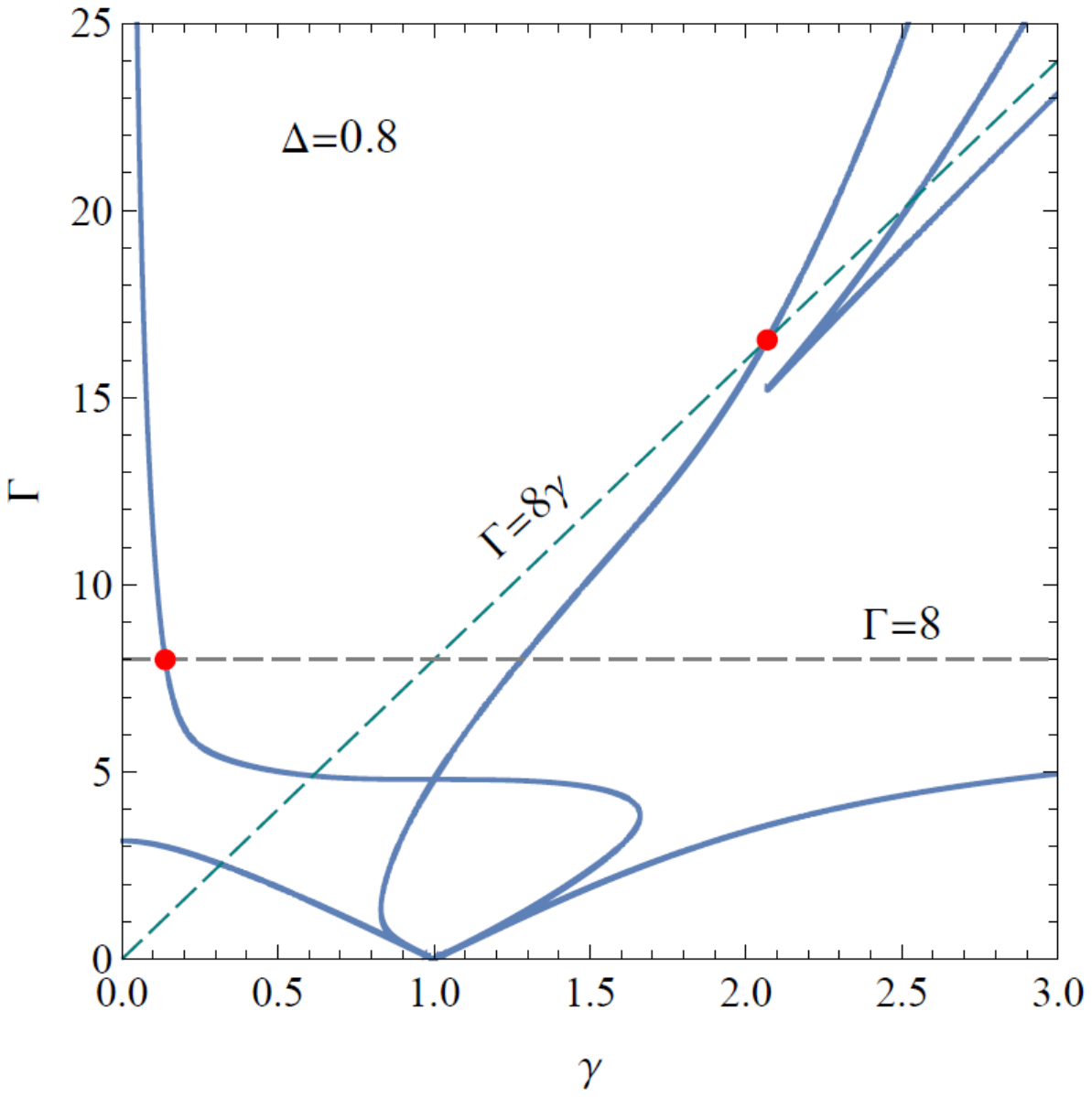}
  \hspace{0.05\columnwidth}
  \includegraphics[width=0.42\columnwidth,clip]{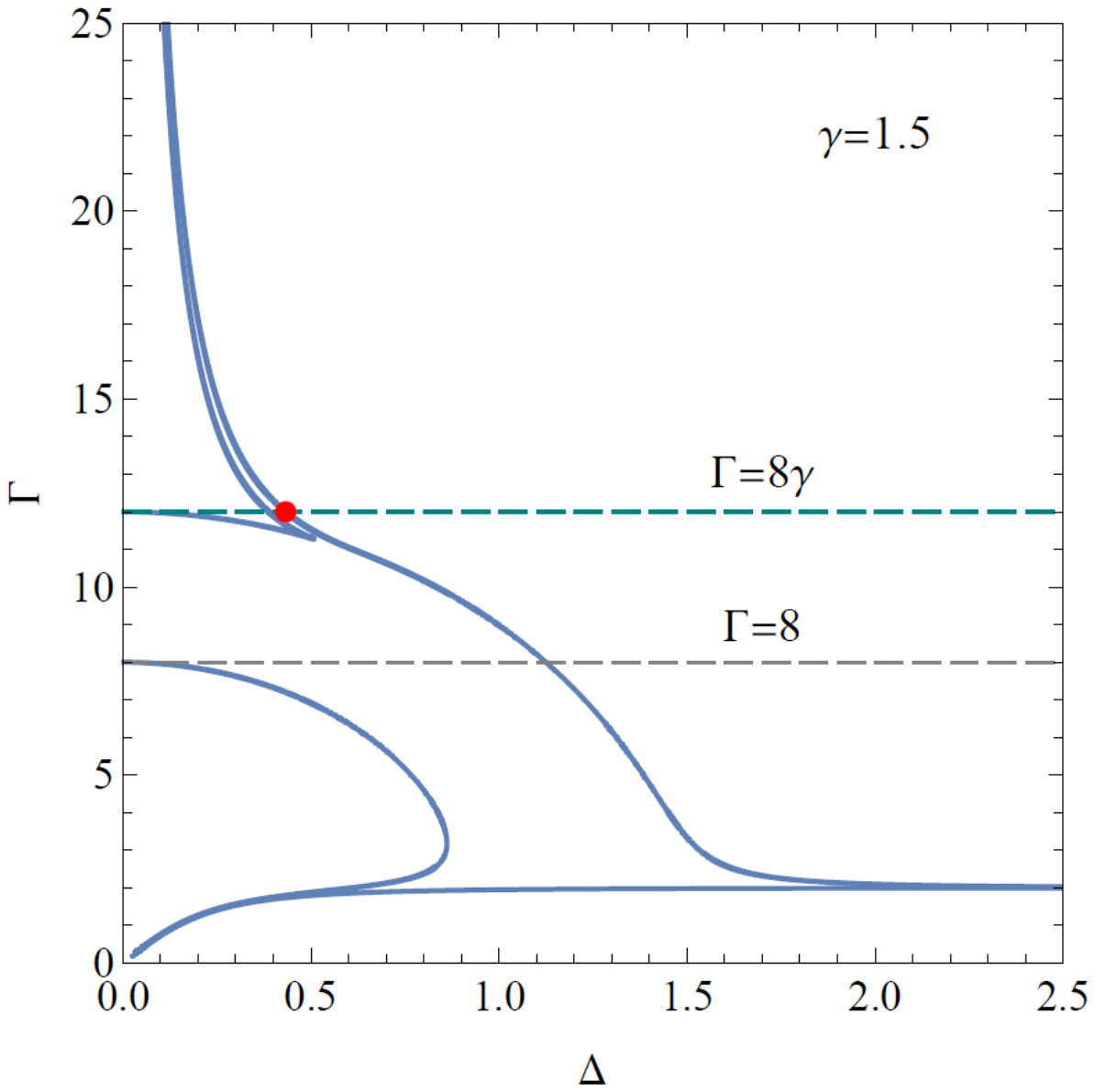}
  \caption{Two-dimensional sections at $\Delta=0.8$ (left panel) and
    $\gamma=1.5$ (right panel) of the manifolds shown in
    Fig.~\ref{Fig-3D}. The dashed straight lines indicate the
    intersections with the $\Gamma=8$ and $\Gamma=8\gamma$ planes (not
    shown in Fig.~\ref{Fig-3D}).}
  \label{Fig-2Dintersection}
\end{figure}

In Fig.~\ref{Fig-2Dintersection} we show the two-dimensional sections
at $\Delta=0.8$ (left panel) and $\gamma=1.5$ (right panel) of the
three-dimensional LEP manifolds reported in Fig.~\ref{Fig-3D}. The
solid blue lines are the sections of the $\Delta$-dependent LEPs
arising from $\Sigma_-$ while the dashed lines indicates the LEPs
$\Gamma=8$ and $\Gamma=8\ga$ from the $\Delta$-independent Liouvillian
block $\Sigma_+$.

\begin{figure}[h]
  \centering \includegraphics[width=0.5\columnwidth]{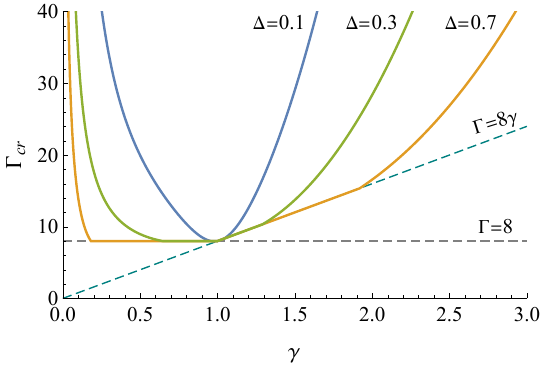}
  \caption{ Two-dimensional sections at $\Delta=0.1$, 03, and 0.7 of
    $\Gamma_{cr}(\gamma,\Delta)$ of Fig.~\ref{Fig-PhaseDiagram}.  The
    dashed lines are the sections of the planes $\Gamma=8$ and
    $\Gamma=8\gamma$. }
  \label{Fig-PhaseDiagramCuts}
\end{figure}

In Fig.~\ref{Fig-PhaseDiagramCuts} we provide the two-dimensional
sections at $\Delta=0.1$, 03, and 0.7 of $\Gamma_{cr}(\gamma,\Delta)$
given in Fig.~\ref{Fig-PhaseDiagram}.  Also in this case, the dashed
lines are the sections of the planes $\Gamma=8$ and $\Gamma=8\gamma$
which partially coincide with the $\Delta$-dependent sections shown by
solid lines.

\section{D.  Bifurcation diagrams at points a,b,c,d in
  Fig.~\ref{Fig-PhaseDiagram} of the main text}
\begin{figure}[h]
  \centering
  \includegraphics[width=0.45\columnwidth,clip]{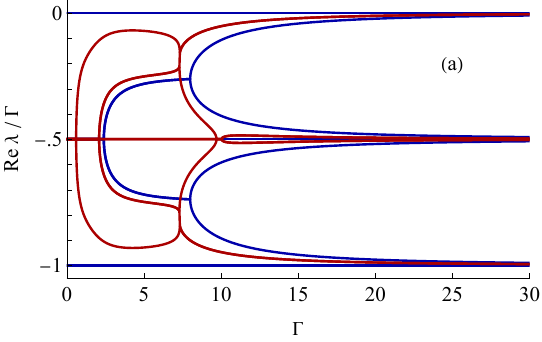}
  \hspace{0.05\columnwidth}
  \includegraphics[width=0.45\columnwidth,clip]{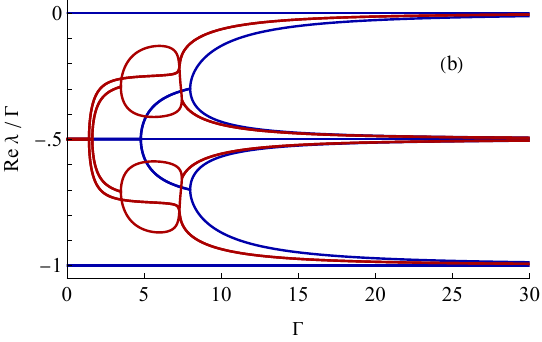}
  \includegraphics[width=0.45\columnwidth,clip]{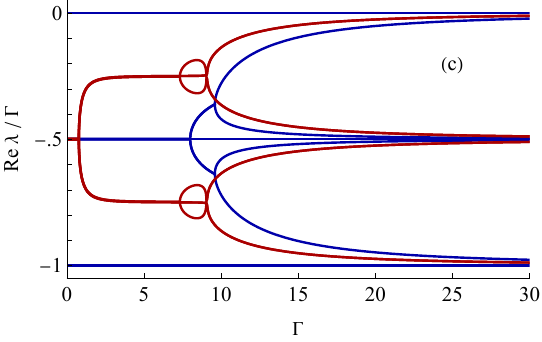}
  \hspace{0.05\columnwidth}
  \includegraphics[width=0.45\columnwidth,clip]{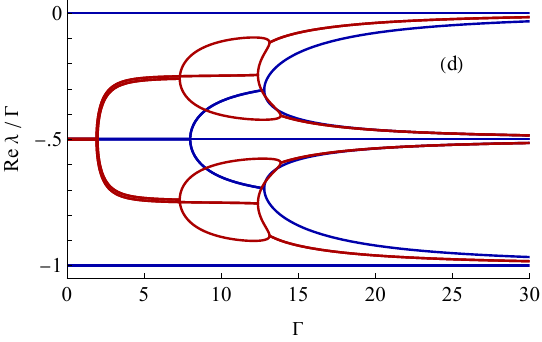}
  \caption{ Rescaled real parts of the Liouvillian eigenvalues versus
    $\Gamma$ for parameter values $(\gamma,\Delta)=(0.3, 0.4)$,
    $(0.6, 0.4)$, $(1.2, 0.4)$, and $(1.6, 0.4)$ corresponding,
    respectively, to the points a,b,c, and d depicted in
    Fig.~\ref{Fig-PhaseDiagram} of the main text.  Blue and red dots
    refer to $\Delta$-independent and $\Delta$-dependent eigenvalues,
    respectively. Notice that for the $\Delta$-independent part of the
    spectrum the first and second left most branching points occur at
    $\Gamma=\min(8, 8\gamma)$ and at $\Gamma=\max(8, 8\gamma)$,
    respectively.  }
  \label{FigSMbifurcationDiagrams}
\end{figure}

In Fig.~\ref{FigSMbifurcationDiagrams} we depict the bifurcation
diagrams obtained for the rescaled real part of all Liouvillian
eigenvalues (from $\Sigma_-$ as well as from $\Sigma_+$) as a function
of $\Gamma$ for parameter values corresponding to points a, b, c, d of
Fig.~\ref{Fig-PhaseDiagram}.  In cases a,d the largest bifurcation
belongs to the $\Delta$-dependent Liouvillian eigenvalues (red
curves), while in cases b,c it belongs to the $\Delta$-independent
Liouvillian eigenvalues (blue curves), in full agreement with the
analysis done in the main text, see Fig.~\ref{Fig-PhaseDiagram}.

\section{E.  Obtaining the boundary lines in
  Fig.~\ref{Fig-PhaseDiagram}}
Here we explain how we obtain the red boundary lines, shown in both
panels of Fig.~\ref{Fig-PhaseDiagram}, separating the regions where
$\Gamma_{cr}$ corresponds to a LEP arising from an eigenvalue of
$\Sigma_+$ or $\Sigma_-$.

To determine the left part of the curve, we solve numerically
Eq.~(\ref{polyLEP}) for $\De$ fixing the value of $\Ga=8$ and varying
$\ga$ in the interval $0<\ga<1$.  This yields two branches of
solutions $\De_{1L}(\ga)>0$ and $\De_{2L}(\ga)>\De_{1L}(\ga) $.  The
upper branch $\De_{2L}y(\ga)$ gives the red line in
Fig.~\ref{Fig-PhaseDiagram} in the interval $0<\ga<1$.

Likewise, in the region $1<\ga$ we solve Eq.~(\ref{polyLEP}) for
$\De$, fixing the value of $\Ga=8\ga$ and varying $\ga>1$.  Also in
this case we obtain two branches of solutions $\De_{1R}(\ga)>0$ and
$\De_{2R}(\ga)>\De_{1R}(\ga) $.  The upper branch $\De_{2R}(\ga)$
gives the red line in Fig.~\ref{Fig-PhaseDiagram} in the region
$\ga>1$.

\section{F. Non-diagonalizability of the Liouvillian on LEPMs}
In this section we show that on the LEPMs the Liouvillian matrix is
non-diagonalizable. The analysis is made on the two diagonal blocks
$\Sigma_\pm$, separately.  For the block $\Sigma_+$ the LEPMs are
$\Delta$ independent, they are the points of the two planes $\Gamma=8$
and $\Gamma=8\gamma$. Eigenvalues and eigenvectors of $\Sigma_+$ at
these points are reported in Table \ref{table1} and \ref{table2},
respectively.  In both cases, we note the presence of the null
eigenvalue in the spectrum, meaning that the NESS of the system,
obtained from the corresponding eigenvector, belongs to the block
$\Sigma_+$.

\squeezetable
\begin{table}
  \renewcommand{\arraystretch}{2.4}
  \begin{ruledtabular}
    \begin{tabular}{cccccccc}
      $-8$
      &$-4$
      &$-4$
      &$0$
      &$2 (-2+\sqrt{1-\gamma ^2})$
      &$2 (-2+\sqrt{1-\gamma ^2})$
      &$-2 (2+\sqrt{1-\gamma ^2})$
      &$-2 (2+\sqrt{1-\gamma^2})$
      \\ \hline
      $-1$
      &$0$
      &$0$
      &$\frac{i (1+\gamma ) (17-2 \gamma +\gamma ^2)}{4 (-1+\gamma )^2}$
      &$-\frac{i (4+4 \gamma +3 \sqrt{1-\gamma ^2}+\gamma\sqrt{1-\gamma ^2})}
        {(1+\gamma ) (2+\sqrt{1-\gamma ^2})}$
      &$0$
      &$-\frac{i (-4-4 \gamma +3 \sqrt{1-\gamma ^2}+\gamma \sqrt{1-\gamma^2})}
        {(1+\gamma)(-2+\sqrt{1-\gamma ^2})}$
      &$0$
      \\
      $-1$
      &$0$
      &$0$
      &$\frac{1}{4} i (1+\gamma )$
      &$\frac{i (-\sqrt{1-\gamma ^2}+\gamma  \sqrt{1-\gamma ^2})}
        {(1+\gamma ) (2+\sqrt{1-\gamma^2})}$
      &$0$
      &$\frac{i (-\sqrt{1-\gamma ^2}+\gamma \sqrt{1-\gamma ^2})}
        {(1+\gamma ) (-2+\sqrt{1-\gamma ^2})}$
      &$0$
      \\
      $0$
      &$0$
      &$1$
      &$-\frac{1+\gamma }{-1+\gamma }$
      &$-\frac{1-\gamma }{\sqrt{1-\gamma ^2}}$
      &$0$
      &$-\frac{-1+\gamma }{\sqrt{1-\gamma ^2}}$
      &$0$
      \\
      $0$
      &$0$
      &$1$
      &$-\frac{-1-\gamma }{-1+\gamma }$
      &$-\frac{-1+\gamma }{\sqrt{1-\gamma ^2}}$
      &$0$
      &$-\frac{1-\gamma }{\sqrt{1-\gamma ^2}}$
      &$0$
      \\
      $1$
      &$0$
      &$0$
      &$\frac{i (17+2 \gamma +\gamma ^2)}{4 (1+\gamma )}$
      &$\frac{2 i \sqrt{1-\gamma ^2}}{1+\gamma }+\frac{i (1+\gamma )}
        {2+\sqrt{1-\gamma^2}}$
      &$0$
      &$\frac{i (-3-2 \gamma +\gamma ^2+4 \sqrt{1-\gamma ^2})}
        {(1+\gamma ) (-2+\sqrt{1-\gamma ^2})}$
      &$0$
      \\
      $1$
      &$0$
      &$0$
      &$\frac{1}{4} i (1+\gamma )$
      &$\frac{i (1+\gamma )}{2+\sqrt{1-\gamma ^2}}$
      &$0$
      &$-\frac{i (1+\gamma )}{-2+\sqrt{1-\gamma ^2}}$
      &$0$
      \\
      $0$
      &$1$
      &$0$
      &$-1$
      &$-1$
      &$0$
      &$-1$
      &$0$
      \\
      $0$
      &$1$
      &$0$
      &$1$
      &$1$
      &$0$
      &$1$
      &$0$
    \end{tabular}
  \end{ruledtabular}
  \caption{Eigenvalues (top row) and eigenvectors (corresponding
    underlying columns) of the $\Sigma_+$ diagonal block of the
    Liouvillian on the LEPM $\{\gamma,\Delta,\Gamma = 8\}$. }
  \label{table1}
\end{table}
\squeezetable
\begin{table}
  \renewcommand{\arraystretch}{2.4}
  \begin{ruledtabular}
    \begin{tabular}{cccccccc}
      $0$
      &$-8 \gamma$
      &$-4 \gamma$
      &$-4 \gamma$
      &$2 (\sqrt{\gamma^2-1}-2 \gamma )$
      &$2 (\sqrt{\gamma^2-1}-2 \gamma )$
      &$-2 (\sqrt{\gamma^2-1}+2 \gamma )$
      &$-2 (\sqrt{\gamma^2-1}+2 \gamma )$
      \\ \hline
      $\frac{i (1+\gamma ) (1-2 \gamma +17 \gamma^2)}
      {4 (-1+\gamma )^2 \gamma }$
      &$-1$
      &$0$
      &$0$
      &$-\frac{i (4 \gamma +4 \gamma^2+\sqrt{-1+\gamma^2}+3 \gamma
        \sqrt{-1+\gamma^2})}{(1+\gamma ) (2 \gamma +
        \sqrt{-1+\gamma^2})}$
      &$0$
      &$-\frac{i (4\gamma +4 \gamma^2-\sqrt{-1+\gamma^2}-3 \gamma
        \sqrt{-1+\gamma^2})}{(1+\gamma )
        (2 \gamma -\sqrt{-1+\gamma^2})}$
      &$0$
      \\
      $\frac{i (1+\gamma )}{4 \gamma }$
      &$-1$
      &$0$
      &$0$
      &$-\frac{i (-\sqrt{-1+\gamma^2}+\gamma\sqrt{-1+\gamma^2})}
        {(1+\gamma )(2 \gamma +\sqrt{-1+\gamma^2})}$
      &$0$
      &$-\frac{i (-\sqrt{-1+\gamma^2}+\gamma \sqrt{-1+\gamma^2})}
        {(1+\gamma ) (-2 \gamma +\sqrt{-1+\gamma^2})}$
      &$0$
      \\
      $-\frac{1+\gamma }{-1+\gamma }$
      &$0$
      &$0$
      &$1$
      &$-\frac{1-\gamma }{\sqrt{-1+\gamma^2}}$
      &$0$
      &$-\frac{-1+\gamma }{\sqrt{-1+\gamma^2}}$
      &$0$
      \\
      $-\frac{-1-\gamma }{-1+\gamma }$
      &$0$
      &$0$
      &$1$
      &$-\frac{-1+\gamma }{\sqrt{-1+\gamma^2}}$
      &$0$
      &$-\frac{1-\gamma }{\sqrt{-1+\gamma^2}}$
      &$0$
      \\
      $\frac{i (1+2 \gamma +17 \gamma^2)}{4 \gamma (1+\gamma )}$
      &$1$
      &$0$
      &$0$
      &$\frac{2 i \sqrt{-1+\gamma^2}}{1+\gamma } + \frac{i (1+\gamma)}
        {2 \gamma +\sqrt{-1+\gamma^2}}$
      &$0$
      &$-\frac{2 i \sqrt{-1+\gamma^2}}{1+\gamma }-\frac{i (1+\gamma )}
        {-2 \gamma +\sqrt{-1+\gamma^2}}$
      &$0$
      \\
      $\frac{i (1+\gamma )}{4 \gamma }$
      &$1$
      &$0$
      &$0$
      &$\frac{i (1+\gamma )}{2 \gamma +\sqrt{-1+\gamma^2}}$
      &$0$
      &$-\frac{i (1+\gamma )}{-2 \gamma +\sqrt{-1+\gamma^2}}$
      &$0$
      \\
      $-1$
      &$0$
      &$1$
      &$0$
      &$-1$
      &$0$
      &$-1$
      &$0$
      \\
      $1$
      &$0$
      &$1$
      &$0$
      &$1$
      &$0$
      &$1$
      &$0$
    \end{tabular}
  \end{ruledtabular}
  \caption{As in Table~\ref{table1} on the LEPM
    $\{\gamma,\Delta,\Gamma = 8\gamma \}$. }
  \label{table2}
\end{table}

From Tables~\ref{table1} and \ref{table2} it is evident that the
coalescence of two pairs of complex eigenvalues (5th-6th and 7th-8th
eigenvalues in both tables) is associated to the appearance of two
null eigenvectors in the corresponding eigenspace.  In fact, the
dimension of the eigenspace is reduced as in the general case of
linearly dependent eigenvectors. In other words, the $\Sigma_+$ matrix
becomes non-diagonalizable and its Jordan decomposition provides the
typical Jordan blocks shown in Table~\ref{table3}.

\begin{table}
  \begin{tabular}{cc}
    $
    \Sigma_+(\gamma,\Delta,\Gamma=8)=
    \begin{pmatrix}
      -8 & 0 & 0 & 0 & 0 & 0 & 0 & 0 \\
      0 & -4 & 0 & 0 & 0 & 0 & 0 & 0 \\
      0 & 0 & -4 & 0 & 0 & 0 & 0 & 0 \\
      0 & 0 & 0 & 0 & 0 & 0 & 0 & 0 \\ \cline{5-6}
      0 & 0 & 0 & 0 & \multicolumn{1}{|c}{\alpha_+}
                         & \multicolumn{1}{c|}{1} & 0 & 0 \\
      0 & 0 & 0 & 0 & \multicolumn{1}{|c}{0}
                         & \multicolumn{1}{c|}{\alpha_+} & 0 & 0 \\
      \cline{5-6}\cline{7-8}
      0 & 0 & 0 & 0 & 0 & 0 & \multicolumn{1}{|c}{\alpha_-}
                                 & \multicolumn{1}{c|}{1} \\
      0 & 0 & 0 & 0 & 0 & 0 & \multicolumn{1}{|c}{0}
                                 & \multicolumn{1}{c|}{\alpha_-}\\ \cline{7-8}
    \end{pmatrix}
    $
    &$\qquad
      \Sigma_+(\gamma,\Delta,\Gamma=8\gamma)=
      \begin{pmatrix}
        0 & 0 & 0 & 0 & 0 & 0 & 0 & 0 \\
        0 & -8 \gamma & 0 & 0 & 0 & 0 & 0 & 0 \\
        0 & 0 & -4 \gamma & 0 & 0 & 0 & 0 & 0 \\
        0 & 0 & 0 & -4 \gamma & 0 & 0 & 0 & 0 \\ \cline{5-6}
        0 & 0 & 0 & 0 & \multicolumn{1}{|c}{\beta_+}
                          & \multicolumn{1}{c|}{1} & 0 & 0 \\
        0 & 0 & 0 & 0 & \multicolumn{1}{|c}{0}
                          & \multicolumn{1}{c|}{\beta_+} & 0 & 0 \\
        \cline{5-6}\cline{7-8}
        0 & 0 & 0 & 0 & 0 & 0 & \multicolumn{1}{|c}{\beta_-}
                                  & \multicolumn{1}{c|}{1} \\
        0 & 0 & 0 & 0 & 0 & 0 & \multicolumn{1}{|c}{0}
                                  & \multicolumn{1}{c|}{\beta_-}\\ \cline{7-8}
      \end{pmatrix}
      $
  \end{tabular}
  \caption{Jordan block form of $\Sigma_+$ on the LEPMs relative to
    Table~\ref{table1} (left matrix) and Table~\ref{table2} (right
    matrix).  We put $\alpha_\pm=-4 \pm 2 \sqrt{1-(8\gamma)^2}$ and
    $\beta_\pm=-4 \gamma \pm 2 \sqrt{\gamma^2-1}$. }
  \label{table3}
\end{table}

Similar results are obtained for the $\Sigma_-$ block of the
Liouvillian. However, in the case of this $\Delta$-dependent block the
LEPMs are much more involved with complicated topology and several
sheets. Taking only positive real values for $\gamma,\Delta$ and
restricting to only real positive solutions of the polynomial equation
(\ref{polyLEP}), one can show that LEPM can have a number of sheets
(branches) that varies from $1$ to $6$ depending on the values of
$\gamma,\Delta$.

Analytical expressions for the Jordan block decomposition of
$\Sigma_-$, except for a few simple cases (see below), are impossible
to derive and one must recourse to numerical calculations. Using
numerical solutions of Eq.~(\ref{polyLEP}) one finds for generic
points on a LEPM, results qualitatively similar to those obtained for
the block $\Sigma_+$ with the difference that the NESS does not belong
to the manifold and the number of pairs of complex coalescing
eigenvalues can be maximal, i.e., as large as $4$, depending on
$\gamma, \Delta$ values.

A particularly simple case in which the coalescence of the eigenvalues
and eigenvectors of the $\Sigma_-$ block can be checked analytically
is obtained for $\gamma=1$.  In this case the coefficients of the
polynomial appearing in Eq.~(\ref{polyLEP}) drastically simplify and
it admits the real positive root $\Gamma=8$ for all values of
$\Delta$. The corresponding LEPM then becomes the EP line
$\{\gamma=1, \Delta, \Gamma=8\}$. Eigenvalues and eigenvectors of the
$\Sigma_-$ block along this line are reported in Table~\ref{table4}
and the corresponding Jordan block decomposition is given in
Table~\ref{table5}.

\squeezetable
\begin{table}
  \renewcommand{\arraystretch}{2.4}
  \begin{ruledtabular}
    \begin{tabular}{cccccccc}
      $-6-2i\De$
      &$-6-2i\De$
      &$-2-2i\De$
      &$-2-2i\De$
      &$-6+2i\De$
      &$-6+2i\De$
      &$-2+2i\De$
      &$-2+2i\De$
      \\ \hline
      $\frac{-2\De+i}{(\De+i)^2}$
      &$0$
      &$0$
      &$0$
      &$0$
      &$0$
      &$i$
      &$0$
      \\
      $0$
      &$0$
      &$0$
      &$0$
      &$i(\De+i)^2$
      &$0$
      &$0$
      &$0$
      \\
      $-\frac{1}{(\De+i)^2}$
      &$0$
      &$0$
      &$0$
      &$0$
      &$0$
      &$1$
      &$0$
      \\
      $0$
      &$0$
      &$0$
      &$0$
      &$-i(\De+i)^2$
      &$0$
      &$0$
      &$0$
      \\
      $0$
      &$0$
      &$-i$
      &$0$
      &$2\De+i$
      &$0$
      &$0$
      &$0$
      \\
      $i$
      &$0$
      &$0$
      &$0$
      &$0$
      &$0$
      &$0$
      &$0$
      \\
      $1$
      &$0$
      &$0$
      &$0$
      &$0$
      &$0$
      &$0$
      &$0$
      \\
      $0$
      &$0$
      &$1$
      &$0$
      &$1$
      &$0$
      &$0$
      &$0$
    \end{tabular}
  \end{ruledtabular}
  \caption{Eigenvalues and eigenvectors of the diagonal block
    $\Sigma_-$, arranged as in Table~\ref{table1}, on the EP line
    $\{\gamma=1,\Delta,\Gamma = 8\}$.  }
  \label{table4}
\end{table}

\begin{table}
  \begin{tabular}{c}
    $
    \Sigma_-(\gamma=1,\Delta,\Gamma=8)=
    \begin{pmatrix}
      \cline{1-2}
      \multicolumn{1}{|c}{-8-2i} & \multicolumn{1}{c|}{1}
      &0 & 0 & 0 & 0 & 0 & 0 \\
      \multicolumn{1}{|c}{0} & \multicolumn{1}{c|}{-8-2i}
      & 0 & 0 & 0 & 0 & 0 & 0 \\ \cline{1-2}\cline{3-4}
      0 & 0 & \multicolumn{1}{|c}{-6+2i} & \multicolumn{1}{c|}{1}
             & 0 & 0 & 0 & 0 \\
      0 & 0 & \multicolumn{1}{|c}{0} & \multicolumn{1}{c|}{-6+2i}
             & 0 & 0 & 0 & 0 \\ \cline{3-4}\cline{5-6}
      0 & 0 & 0 & 0 & \multicolumn{1}{|c}{-4-2i} & \multicolumn{1}{c|}{1}
                     & 0 & 0 \\
      0 & 0 & 0 & 0 & \multicolumn{1}{|c}{0} & \multicolumn{1}{c|}{-4-2i}
                     & 0 & 0 \\ \cline{5-6}\cline{7-8}
      0 & 0 & 0 & 0 & 0 & 0 & \multicolumn{1}{|c}{-2+2i}
                         & \multicolumn{1}{c|}{1} \\
      0 & 0 & 0 & 0 & 0 & 0 & \multicolumn{1}{|c}{0}
                         & \multicolumn{1}{c|}{-2+2i} \\ \cline{7-8}
    \end{pmatrix}
    $
  \end{tabular}
  \caption{Jordan block form of $\Sigma_-$ on the LEPMs
    $\{\gamma=1, \Delta, \Gamma=8\}$. }
  \label{table5}
\end{table}

\begin{figure}
  \centerline{\includegraphics[width=0.30\columnwidth,clip]{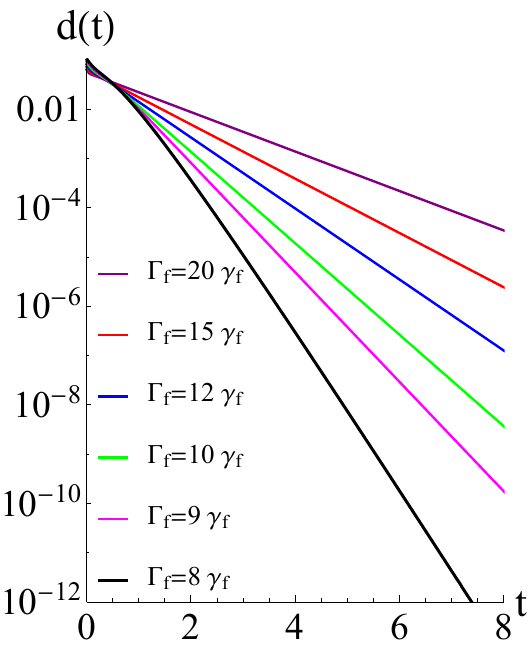}
    \includegraphics[width=0.30\columnwidth,clip]{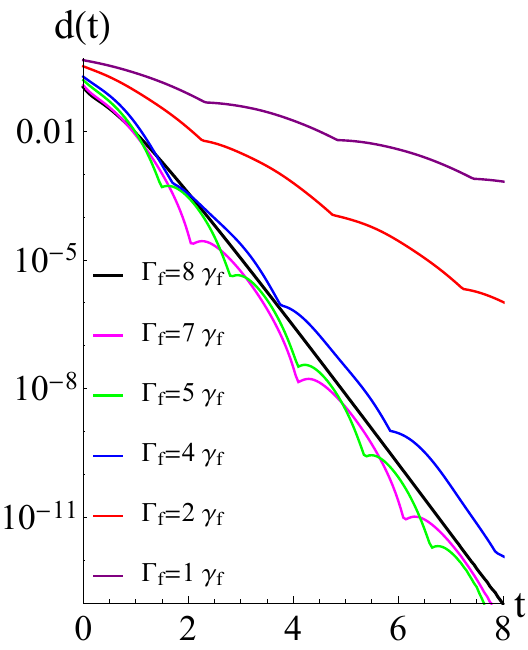} }
  \caption{\rev{Distance
      $d(t)\equiv {\|\rho(t)-\rho_{\text{NESS}}(\gamma_f)\|}_2$ versus
      time $t$ for rapid quenches in the parameter $\gamma$, from
      $\gamma_{i}=1.1$ to $\gamma_{f}=1.6$, on different
      $\Gamma=n \gamma$ planes as indicated in the legend, for
      arbitrary $\De$. The left panel shows data for planes above the
      $\Gamma=8 \gamma$ LEPM, while the right panel data for planes
      below it.}}
  \label{8gamma-relax}
\end{figure}

\rev{
  \section{G. Relaxation times near and on the $\Gamma=8\gamma$
    LEPM.}%
  In the following, we show that in the region $\gamma>1$ of the
  parameter space that characterizes the effective Zeno regime with
  $\Gamma_{cr}=8\gamma$, the fastest relaxation from an initial
  perturbed state to the exact NESS is achieved on the
  $\Gamma=8 \gamma$ LEPM. In this case, the quench involves two
  parameters and is done from ${\gamma_{i}, \Gamma_{i} = 8\gamma_{i}}$
  to ${\gamma_{f}, \Gamma_{f} = 8\gamma_{f}}$. The results obtained
  are depicted in Fig.~\ref{8gamma-relax} from which we see that the
  fastest relaxation is achieved on the $\Gamma=8 \gamma$ plane.  }

\section{H.  Convergence to NESS and characteristic dissipation value
  $\Ga_{ch}$.}
\label{AppZeno}

The exact NESS of the two-qubit problem (when it is unique) can be
evaluated analytically and is given by
\begin{align}
  \rho_\mathrm{NESS}(\Ga)=\left(
  \begin{array}{cccc}
    \frac{(\gamma +1)^2 \left(4 \gamma^2-8 \gamma +\Gamma^2+4\right)}
    {2 \left(8 \gamma^4+\gamma^2 \left(\Gamma^2-16\right)+\Gamma^2+8\right)}
    & 0
    & 0
    & -\frac{i(\gamma -1) (\gamma +1)^2 \Gamma }
      {8 \gamma^4+\gamma^2 \left(\Gamma^2-16\right)+\Gamma^2+8}
    \\
    0
    & \frac{(\gamma -1)^2 \left(4 \gamma^2+8 \gamma +\Gamma^2+4\right)}
      {2 \left(8\gamma^4+\gamma^2 \left(\Gamma^2-16\right)+\Gamma^2+8\right)}
    & \frac{i (\gamma-1)^2 (\gamma +1) \Gamma }
      {8 \gamma^4+\gamma^2 \left(\Gamma^2-16\right)+\Gamma^2+8}
    & 0
    \\
    0
    & -\frac{i (\gamma -1)^2 (\gamma +1) \Gamma }
      {8 \gamma^4+\gamma^2 \left(\Gamma^2-16\right)+\Gamma^2+8}
    & \frac{2 (\gamma -1)^2 (\gamma +1)^2}
      {8 \gamma^4+\gamma^2 \left(\Gamma^2-16\right)+\Gamma^2+8}
    & 0
    \\
    \frac{i (\gamma -1) (\gamma +1)^2 \Gamma }
    {8 \gamma^4+\gamma^2 \left(\Gamma^2-16\right)+\Gamma^2+8}
    & 0
    & 0
    & \frac{2 (\gamma -1)^2 (\gamma +1)^2}
      {8 \gamma^4+\gamma^2 \left(\Gamma^2-16\right)+\Gamma^2+8}
  \end{array}
  \right).
  \label{rhoness}
\end{align}
Note that the NESS is $\De$-independent. In the quantum Zeno limit we
have
\begin{align}
  \rho_\mathrm{Zeno} = \lim_{\Ga \rightarrow \infty} \rho_\mathrm{NESS}(\Ga)
  = \ket{\uparrow} \bra{\uparrow}\otimes
  \left(
  \begin{array}{cc}
    \frac{(\gamma +1)^2}{2 \left(\gamma^2+1\right)} & 0 \\
    0 & \frac{(\gamma -1)^2}{2 \left(\gamma^2+1\right)} \\
  \end{array}
  \right),
\end{align}
in accordance with Eq.~(\ref{eq:ZenoNESS}) obtained in the main text
with the help of the reduced Zeno dynamics \cite{2018ZenoDynamics}.
We observe that for large $\Ga$
\begin{align}
  \tr(\rho_\mathrm{Zeno}^2) - \tr(\rho_\mathrm{NESS}^2(\Ga))
  = O\left( \frac{1}{\Ga^2}\right) + \dots
  \label{AppDifferenceOfTraces}
\end{align}
so that this quantity can serve as a measure of the distance to the
Zeno NESS for fixed values of the dissipation $\Gamma$.

Using Eq.~(\ref{AppDifferenceOfTraces}) we introduce a characteristic
dissipation strength $\Ga_{ch}$ needed to reach the Zeno NESS, as
\begin{align}
  \Ga_{ch}^2(\ga) = \lim_{\Ga \rightarrow \infty } \Ga^2
  \left[ \tr(\rho_\mathrm{Zeno}^2)- \tr(\rho_\mathrm{NESS}^2(\Ga)) \right]
  = \frac{4 \gamma^{10}+36 \gamma^8-40 \gamma^6-40 \gamma^4+36 \gamma^2+4}
  {\gamma^8+4\gamma^6+6 \gamma^4+4 \gamma^2+1}.
  \label{Ga-char}
\end{align}
First of all, note that $\Ga_{ch} $ is independent of $\De$ and also
it does not have any singularities for $\ga \rightarrow 0$. This might
seem in contradiction to what is stated in
Eq.~(\ref{eq:GaDivergencies}). To resolve the issue, we remark that
the effective dynamics (\ref{LMEeff}) only concerns the relaxation of
the diagonal elements of the reduced density matrix, i.e., those
elements which in the limit $\Gamma\to\infty$ become the NESS
eigenvalues. Expression (\ref{Ga-char}) thus gives an estimate of the
relaxation of a part of the system only, while for the relaxation of
the non-diagonal part of the reduced density matrix $\rho$ the full
system (\ref{LME}) still needs to be considered.


\begin{thebibliography}{99}
\bibitem{note_vec} Given the matrix $A=[A_{ij}]_{i,j=1}^{n}$ we define
  the column vector
  $\vec{A}=\transpose{(A_{11},\dots,A_{1n},\dots,A_{n1},\dots,A_{nn})}$,
  where $\transpose{(\cdot)}$ indicates matrix transpose. This implies
  $\vec{A B} = (A \otimes I_n) \vec{B} = (\transpose{B} \otimes I_n)
  \vec{A}$ and $\vec{A B C} = (A \otimes \transpose{C}) \vec{B}$, with
  $A,B,C$ arbitrary $n\times n$ matrices.
\bibitem{nota_reali} We restrict to real parameters to keep contact
  with realistic physical settings, however, LEPMs exist for complex
  values of parameters as well.
\end{thebibliography}
\end{document}